\documentclass[12pt,a4paper]{article}%
\usepackage{epsfig}
\usepackage{amsmath}
\usepackage{amsfonts}
\usepackage{amssymb}
\usepackage{graphicx}
\usepackage[T1]{fontenc}
\usepackage[utf8]{inputenc}
\usepackage{authblk}%
\providecommand{\U}[1]{\protect\rule{.1in}{.1in}}
\begin{document}

\title{Comment on: "Quantum dynamics of a general time-dependent coupled oscillator".}
\author{R. Zerimeche$^{a,b}$, N. Mana$^{a}$\thanks{E-mail: na3ima\_mn@hotmail.fr} , M.
Sekhri$^{a}$, N. Amaouche$^{a}$ and M. Maamache$^{a}$\\$^{a}$Laboratoire de Physique Quantique et Syst\`{e}mes Dynamiques,\\Facult\'{e} des Sciences, Universit\'{e} Ferhat Abbas S\'{e}tif 1, S\'{e}tif
19000, Algeria.\\$^{b\text{ \ }}$Physics Department, University of Jijel, BP 98, Ouled Aissa,
18000 Jijel, Algeria{\small .}}
\date{}
\maketitle

\begin{abstract}
By using dynamical invariants theory, Hassoul et al. \cite{Hassoul, Hassoul2}
investigate the quantum dynamics of two (2D) and three (3D) dimensional
time-dependent coupled oscillators. They claim that, in the 2D case,
introducing two pairs of annihilation and creation operators uncouples the
original invariant operator so that it becomes the one that describes two
independent subsystems. For the 3D case, the authors pretend that they have
obtained a diagonalized invariant which is exactly the sum of three simple
harmonic oscillators.

We show that their investigations suffer from basic errors and therefore the
found results are not valid .

\end{abstract}

\section{Introduction}

The harmonic oscillator is one of the most important models in quantum
mechanics, and one of the few ones that has an exact analytical solution what
made it applicable in the study of the dynamical properties of\ different
physical systems. Recently, Hassoul et al \cite{Hassoul, Hassoul2} have
investigated quantum dynamical properties of a two (2D) and three (3D)
dimensional time-dependent coupled oscillator. Their studies are based on the
theory of two-dimensional (2D) and three-dimensional (3D) dynamical invariants.

Oscillators are the core of both natural systems and technological devices
which made the study of dynamical properties of coupled oscillatory systems
extremely demanded though their complex motion especially when the parameters
are time-dependent and/or the dimension of the system is more than two.

The time-dependent coupled oscillator that has been a point of interest in the
research field for a few years now \cite{1}-\cite{6}, modele various physical
systems \cite{7}-\cite{17} and helped in explaining numerous physical
interacting systems including trapped atoms \cite{1'}, nano-optomechanical
resonances \cite{6', 7'}, electromagnetically induced transparency \cite{8'},
stimulated Raman effects \cite{9'}, time-dependent Josephson phenomena
\cite{10'}, and systems of three isotropically coupled spins 1/2 \cite{11'}.
Coupled oscillator are fundamental for quantum technologies such as quantum
computing and quantum cryptography \cite{12', 13', 14'}

In what follows, we highlight many basic and mathematical flaws made by
Hassoul et al. in their recent papers \cite{Hassoul, Hassoul2} while
investigating the quantum dynamical properties of a general time-dependent
coupled oscillator using two and three-dimensional dynamical invariants. From
there, we proceeded to reevaluate the entire study and approach the whole
subject in a more scientifically coherent manner. In section 2, we
recapitulate the basic errors made in \cite{Hassoul} where a 2D quantum
invariant operator is introduced for the study of a 2D time-dependent coupled
oscillator system, then we show that the linear canonical transformation
defined in \cite{Hassoul} can not be adapted to the considered problem without
a constraint on the mass and to solve the problem in question, we\textrm{
}introduce more adapting canonical transformations. In Section 3, we consider
the 3D case investigated in \cite{Hassoul2} and highlight with some simple
mathematical calculation all the flawed results found in \cite{Hassoul2}.

\section{Two-dimensional time-dependent harmonic oscillators}

The Hamiltonian of the time-dependent coupled oscillator that Hassoul et al.
\cite{Hassoul} consider is
\begin{equation}
H(t)=\frac{1}{2}\overset{2}{\underset{i=1}{\sum}}\left[  \frac{P_{i}^{2}%
}{m_{i}\left(  t\right)  }+c_{i}(t)X_{i}^{2}\right]  +\frac{1}{2}c_{3}%
(t)X_{1}X_{2}, \label{hamilton}%
\end{equation}
where $m_{i}\left(  t\right)  ,$ $c_{i}(t)$\ and $c_{3}(t)$ are arbitary
functions of time. They assume a quantum invariant operator of the system of
the form%
\begin{equation}
I\left(  t\right)  =\frac{1}{2}\overset{2}{\underset{i=1}{\sum}}\left[
\alpha_{i}\left(  t\right)  P_{i}^{2}+\beta_{i}\left(  t\right)  \left(
X_{i}P_{i}+P_{i}X_{i}\right)  +\gamma_{i}\left(  t\right)  X_{i}^{2}\right]
+\frac{1}{2}\eta\left(  t\right)  X_{1}X_{2}, \label{invar}%
\end{equation}
where the parameters $\alpha_{i}\left(  t\right)  ,$ $\beta_{i}\left(
t\right)  ,$ $\gamma_{i}\left(  t\right)  $ $\left(  i=1,2\right)  $ and
$\eta\left(  t\right)  $ are real and differentiable functions of time. The
substitution of (\ref{hamilton}) and (\ref{invar}) into the invariance
condition%
\begin{equation}
\frac{dI}{dt}=\frac{\partial I}{\partial t}+\frac{1}{i\hbar}\left[
I,H\right]  =0, \label{inv}%
\end{equation}
implies the auxiliary equations given in \cite{Hassoul} as Eqs. (8-14) and an
additional condition which was not mentioned
\begin{equation}
\alpha_{1}\left(  t\right)  m_{1}\left(  t\right)  =\alpha_{2}\left(
t\right)  m_{2}\left(  t\right)  , \label{condit5}%
\end{equation}
we note that the commutator in relation (\ref{inv}) is replaced with Poisson
brackets in the classical case.

The authors of Ref. \cite{Hassoul} pretend to obtain the solution of the
equations (8-13), omitting to mention that the so-called solution
$m_{i}\left(  t\right)  =$ $1/\alpha_{i}\left(  t\right)  $ must obey the
following constraint equation
\begin{equation}
\ddot{m}_{i}(t)-\frac{1}{2}\frac{\dot{m}_{i}^{2}(t)}{m_{i}(t)}+2\left(
\delta_{i}m_{i}(t)-c_{i}\left(  t\right)  \right)  =0, \label{mass}%
\end{equation}
which is difficult to be solved. This last equation (\ref{mass}) is obtained
from the auxiliary equations (8-13)\ by noting that
\begin{equation}
\alpha_{i}(t)\gamma_{i}(t)-\beta_{i}^{2}(t)=\delta_{i}, \label{1}%
\end{equation}
with $\delta_{i}$\ being a real constant. Condition (\ref{1}) is not mentioned
in \cite{Hassoul}. Note that putting $\frac{1}{\rho_{i}^{2}}=m_{i}(t)$ and
changing $c_{i}\left(  t\right)  $ by $m_{i}(t)c_{i}\left(  t\right)  $ gives
the famous auxiliary equation of $\rho_{i}$ \cite{LR, Hartley, Maamache,
Pedrosa}.

We emphasize that the invariant operator (23) in \cite{Hassoul} can not be
decoupled into (35) simply because the canonical transformations (41-42) can
be adapted to the considered problem if and only if $m_{1}(t)=m_{2}(t)$ and
$\omega_{1}^{2}(t)=\omega_{2}^{2}(t)$. However, in order to solve the problem
in question, we the introduce the following more adapting canonical
transformations that should be used as%
\begin{equation}
\left(
\begin{array}
[c]{c}%
X_{1}\\
X_{2}%
\end{array}
\right)  =\left(
\begin{array}
[c]{cc}%
\frac{1}{\sqrt{m_{1}\left(  t\right)  }}\cos\left(  \frac{\theta}{2}\right)  &
-\frac{1}{\sqrt{m_{1}\left(  t\right)  }}\sin\left(  \frac{\theta}{2}\right)
\\
\frac{1}{\sqrt{m_{2}\left(  t\right)  }}\sin\left(  \frac{\theta}{2}\right)  &
\frac{1}{\sqrt{m_{2}\left(  t\right)  }}\cos\left(  \frac{\theta}{2}\right)
\end{array}
\right)  \left(
\begin{array}
[c]{c}%
Q_{1}\\
Q_{2}%
\end{array}
\right)  , \label{transfor}%
\end{equation}
and set%
\begin{equation}
\frac{1}{\sqrt{m_{i}\left(  t\right)  }}P_{i}+\sqrt{m_{i}\left(  t\right)
}\beta_{i}X_{i}=\mathcal{P}_{i}\text{ ,}%
\end{equation}
to write the invariant operator (\ref{invar}) in the form%
\begin{equation}
I(t)=\frac{1}{2}\overset{2}{\underset{i=1}{\sum}}\left(  \mathcal{P}_{i}%
^{2}+\omega_{i}^{2}Q_{i}^{2}\right)  +\Gamma(t)Q_{1}Q_{2}, \label{coupled}%
\end{equation}
this invariant describes a coupled harmonic oscillator where
\begin{align}
\omega_{1}^{2}(t)  &  =\left(  \frac{\int_{0}^{t}\left[  c_{1}\dot{m}%
_{1}/m_{1}\right]  dt}{m_{1}}-\frac{\dot{m}_{1}^{2}}{4m_{1}^{2}}\right)
\cos^{2}\left(  \frac{\theta}{2}\right)  +\left(  \frac{\int_{0}^{t}\left[
c_{2}\dot{m}_{2}/m_{2}\right]  dt}{m_{2}}-\frac{\dot{m}_{2}^{2}}{4m_{2}^{2}%
}\right)  \sin^{2}\left(  \frac{\theta}{2}\right) \nonumber\\
&  +\left(  \frac{\int_{0}^{t}c_{3}\left[  \dot{m}_{1}/2m_{1}+\dot{m}%
_{2}/2m_{2}\right]  dt}{2\sqrt{m_{1}m_{2}}}\right)  \sin\left(  \theta\right)
\text{,}%
\end{align}%
\begin{align}
\omega_{2}^{2}(t)  &  =\left(  \frac{\int_{0}^{t}\left[  c_{1}\dot{m}%
_{1}/m_{1}\right]  dt}{m_{1}}-\frac{\dot{m}_{1}^{2}}{4m_{1}^{2}}\right)
\sin^{2}\left(  \frac{\theta}{2}\right)  +\left(  \frac{\int_{0}^{t}\left[
c_{2}\dot{m}_{2}/m_{2}\right]  dt}{m_{2}}-\frac{\dot{m}_{2}^{2}}{4m_{2}^{2}%
}\right)  \cos^{2}\left(  \frac{\theta}{2}\right) \nonumber\\
&  -\left(  \frac{\int_{0}^{t}c_{3}\left[  \dot{m}_{1}/2m_{1}+\dot{m}%
_{2}/2m_{2}\right]  dt}{2\sqrt{m_{1}m_{2}}}\right)  \sin\left(  \theta\right)
\text{,}%
\end{align}%
\begin{align}
\Gamma(t)  &  =-\left[  \left(  \frac{\int_{0}^{t}\left[  c_{1}\dot{m}%
_{1}/m_{1}\right]  dt}{m_{1}}-\frac{\dot{m}_{1}^{2}}{4m_{1}^{2}}\right)
-\left(  \frac{\int_{0}^{t}\left[  c_{2}\dot{m}_{2}/m_{2}\right]  dt}{m_{2}%
}-\frac{\dot{m}_{2}^{2}}{4m_{2}^{2}}\right)  \right]  \sin\left(
\theta\right) \nonumber\\
&  +\frac{\int_{0}^{t}c_{3}\left[  \dot{m}_{1}/2m_{1}+\dot{m}_{2}%
/2m_{2}\right]  dt}{\sqrt{m_{1}m_{2}}}\cos\left(  \theta\right)  ,
\end{align}
the separation of variables in eq. (\ref{coupled}) requires the coefficient
$\Gamma(t)$ to be equal to zero allowing us to determine the angle $\theta$ as%
\begin{align}
\tan\left(  \theta\right)   &  =\frac{1}{\sqrt{m_{1}m_{2}}}\int_{0}^{t}%
c_{3}\left[  \dot{m}_{1}/2m_{1}+\dot{m}_{2}/2m_{2}\right]  dt\nonumber\\
&  \times\left[  \left(  \frac{1}{m_{1}}\int_{0}^{t}\left[  c_{1}\dot{m}%
_{1}/m_{1}\right]  dt-\frac{\dot{m}_{1}^{2}}{4m_{1}^{2}}\right)  -\left(
\frac{1}{m_{2}}\int_{0}^{t}\left[  c_{2}\dot{m}_{2}/m_{2}\right]
dt-\frac{\dot{m}_{2}^{2}}{4m_{2}^{2}}\right)  \right]  ^{-1},\text{ \ \ }%
\end{align}
this last equation is the same as (32) in \cite{Hassoul} but without
clarification in terms of how it is deduced.

It is useful that we reexpress the invariant operator (\ref{coupled}) in terms
of the annihilation and creation operators $a_{i}$ and $a_{i}^{+},$
respectively, as%
\begin{equation}
I(t)=\sum\limits_{i=1}^{2}\hbar\omega_{i}\left(  a_{i}^{+}a_{i}+\frac{1}%
{2}\right)  ,
\end{equation}
where%
\begin{align}
a_{1}  &  =\frac{1}{\sqrt{2\hbar\omega_{1}}}\left[  \omega_{1}\left(
\sqrt{m_{1}}\cos\left(  \frac{\theta(t)}{2}\right)  X_{1}+\sqrt{m_{2}}%
\sin\left(  \frac{\theta(t)}{2}\right)  X_{2}\right)  \right. \nonumber\\
&  \left.  +i\left(  \frac{\mathcal{P}_{1}}{\sqrt{m_{1}}}\cos\left(
\frac{\theta\left(  t\right)  }{2}\right)  +\frac{\mathcal{P}_{2}}{\sqrt
{m_{2}}}\sin\left(  \frac{\theta(t)}{2}\right)  \right)  \right]  , \label{a1}%
\end{align}%
\begin{align}
a_{1}^{\dagger}  &  =\frac{1}{\sqrt{2\hbar\omega_{1}}}\left[  \omega
_{1}\left(  \sqrt{m_{1}}\cos\left(  \frac{\theta(t)}{2}\right)  X_{1}%
+\sqrt{m_{2}}\sin\left(  \frac{\theta(t)}{2}\right)  X_{2}\right)  \right.
\nonumber\\
&  \left.  -i\left(  \frac{\mathcal{P}_{1}}{\sqrt{m_{1}}}\cos\left(
\frac{\theta\left(  t\right)  }{2}\right)  +\frac{\mathcal{P}_{2}}{\sqrt
{m_{2}}}\sin\left(  \frac{\theta(t)}{2}\right)  \right)  \right]  ,
\label{a1+}%
\end{align}%
\begin{align}
a_{2}  &  =\frac{1}{\sqrt{2\hbar\omega_{2}}}\left[  \omega_{2}\left(
-\sqrt{m_{1}}\sin\left(  \frac{\theta(t)}{2}\right)  X_{1}+\sqrt{m_{2}}%
\cos\left(  \frac{\theta(t)}{2}\right)  X_{2}\right)  \right. \nonumber\\
&  \left.  +i\left(  -\frac{\mathcal{P}_{1}}{\sqrt{m_{1}}}\sin\left(
\frac{\theta\left(  t\right)  }{2}\right)  +\frac{\mathcal{P}_{2}}{\sqrt
{m_{2}}}\cos\left(  \frac{\theta(t)}{2}\right)  \right)  \right]  , \label{a2}%
\end{align}%
\begin{align}
a_{2}^{\dagger}  &  =\frac{1}{\sqrt{2\hbar\omega_{2}}}\left[  \omega
_{2}\left(  -\sqrt{m_{1}}\sin\left(  \frac{\theta(t)}{2}\right)  X_{1}%
+\sqrt{m_{2}}\cos\left(  \frac{\theta(t)}{2}\right)  X_{2}\right)  \right.
\nonumber\\
&  \left.  -i\left(  -\frac{\mathcal{P}_{1}}{\sqrt{m_{1}}}\sin\left(
\frac{\theta\left(  t\right)  }{2}\right)  +\frac{\mathcal{P}_{2}}{\sqrt
{m_{2}}}\cos\left(  \frac{\theta(t)}{2}\right)  \right)  \right]  .
\label{a2+}%
\end{align}

Moreover, to our knowledge, the invariant operator in the Lewis and Riesenfeld
theory \cite{LR} has time-independent eigenvalues whereas the frequencies
$\omega_{i}^{2}$ are time-dependent which does not imply time-independent
eigenvalues of the invariant as claimed in \cite{Hassoul} and consequently it
is not easy to obtain the phases.

\section{Three-dimensional time-dependent harmonic oscillators}

A generalization of the 2D coupled oscillator to a 3D one has been incorrectly
made by Hassoul et al. in \cite{Hassoul2} where the authors study general
time-dependent three coupled nano-optomechanical oscillators. The Hamiltonian
operator of the 3D system reads%
\begin{align}
H(t)  &  =\frac{1}{2}\overset{3}{\underset{i=1}{\sum}}\left[  \frac{P_{i}^{2}%
}{m_{i}\left(  t\right)  }+m_{i}\left(  t\right)  \omega_{i}^{2}\left(
t\right)  X_{i}^{2}\right] \nonumber\\
&  +\frac{1}{2}\left[  k_{12}\left(  t\right)  X_{1}X_{2}+k_{13}\left(
t\right)  X_{1}X_{3}+k_{23}\left(  t\right)  X_{2}X_{3}\right]  , \label{Ham}%
\end{align}
where {}$X_{i}$ and $P_{i}$ are the canonical coordinates and momenta,
$k_{12}\left(  t\right)  ,$ $k_{13}\left(  t\right)  $ and $k_{23}\left(
t\right)  $ are the coupling parameters. They choose, in this case, an
invariant operator of the form%
\begin{align}
O(t)  &  =\frac{1}{2}\overset{3}{\underset{i=1}{\sum}}\left[  A_{i}\left(
t\right)  P_{i}^{2}+B_{i}\left(  t\right)  \left(  X_{i}P_{i}+X_{i}%
P_{i}\right)  +C_{i}\left(  t\right)  X_{i}^{2}\right] \nonumber\\
&  +\frac{1}{2}\left[  D_{12}\left(  t\right)  X_{1}X_{2}+D_{13}\left(
t\right)  X_{1}X_{3}+D_{23}\left(  t\right)  X_{2}X_{3}\right]  , \label{o(t)}%
\end{align}
where the parameters $A_{i}(t),$ $B_{i}\left(  t\right)  ,$\ $C_{i}\left(
t\right)  ,$\ $D_{12}\left(  t\right)  ,$\ $D_{13}\left(  t\right)  $\ and
$D_{23}\left(  t\right)  $\ are time-dependent real and differentiable functions.

They substitute the Hamiltonian and the invariant formulae in the invariance
condition (\ref{inv}) to derive the formulae of the parameters in (\ref{Ham})
where they give six differential equations (equations (8)-(13) in
\cite{Hassoul2}) with their possible solutions (equations (14)-(19). In fact,
they failed again in mentioning other conditions just like before, since the
substitution of (\ref{Ham}) and (\ref{o(t)}) in (\ref{inv}) implies nine
equations given as%

\begin{align}
\dot{A}_{i}\left(  t\right)   &  =-\frac{2B_{i}(t)}{m_{i}\left(  t\right)
},\label{a}\\
\dot{B}_{i}\left(  t\right)   &  =-\frac{C_{i}\left(  t\right)  }{m_{i}\left(
t\right)  }+m_{i}\left(  t\right)  \omega_{i}^{2}\left(  t\right)
A_{i}(t),\label{b}\\
\dot{C}_{i}\left(  t\right)   &  =2m_{i}\left(  t\right)  \omega_{i}%
^{2}\left(  t\right)  B_{i}(t),\label{c}\\
\dot{D}_{12}\left(  t\right)   &  =\frac{k_{12}\left(  t\right)  }{2}\left[
B_{1}\left(  t\right)  +B_{2}\left(  t\right)  \right]  ,\\
\dot{D}_{13}\left(  t\right)   &  =\frac{k_{13}\left(  t\right)  }{2}\left[
B_{1}\left(  t\right)  +B_{3}\left(  t\right)  \right]  ,\\
\dot{D}_{23}\left(  t\right)   &  =\frac{k_{23}\left(  t\right)  }{2}\left[
B_{2}\left(  t\right)  +B_{3}\left(  t\right)  \right]  ,\label{6}\\
\frac{D_{13}\left(  t\right)  }{D_{12}\left(  t\right)  }  &  =\frac
{k_{13}\left(  t\right)  }{k_{12}\left(  t\right)  },\\
\frac{D_{12}\left(  t\right)  }{D_{23}\left(  t\right)  }  &  =\frac
{k_{12}\left(  t\right)  }{k_{23}\left(  t\right)  },\\
\frac{D_{23}\left(  t\right)  }{D_{13}\left(  t\right)  }  &  =\frac
{k_{23}\left(  t\right)  }{k_{13}\left(  t\right)  },
\end{align}

Similarly to the 2D case, the authors of Ref. \cite{Hassoul2} pretend to
obtain the solution of equations (\ref{a})-(\ref{6}), omitting to mention an
important detail: the following constraint equation that the mass must obey
when considering $m_{i}\left(  t\right)  =$ $1/\alpha_{i}\left(  t\right)  $
\begin{equation}
\ddot{m}_{i}(t)-\frac{1}{2}\frac{\dot{m}_{i}^{2}(t)}{m_{i}(t)}+2\left(
\delta_{i}-\omega_{i}^{2}(t)\right)  m_{i}(t)=0, \label{mass'}%
\end{equation}
which is difficult to be solved. This last equation (\ref{mass'}) is obtained
from the auxiliary equations (\ref{a})-(\ref{6})\ by noting that
\begin{equation}
A_{i}(t)C_{i}(t)-B_{i}^{2}(t)=\delta_{i}, \label{delta}%
\end{equation}
with $\delta_{i}$\ being a real constant. Condition (\ref{delta}) is not
mentioned in \cite{Hassoul2}. Note that putting $\frac{1}{\rho_{i}^{2}}%
=m_{i}(t)$ gives the famous auxiliary equation of $\rho_{i}$ \cite{LR,
Hartley, Maamache, Pedrosa}.

Thus the given solutions (equations (14)-(19) in \cite{Hassoul2}) impose a
constraint on the system that the authors did not pay attention to. The system
can not be resolved for any given mass.

The authors proceed, with the aim to have a diagonalized invariant operator
$\mathcal{O}(t),$ to diagonalize the matrix $\Bbbk$ (formula (30) in
\cite{Hassoul2}) using the invertible matrix $%
\mathbb{R}
$ (formula (46) in \cite{Hassoul2}) as%
\begin{equation}%
\mathbb{R}
^{-1}\Bbbk%
\mathbb{R}
=\left(
\begin{array}
[c]{ccc}%
M_{11} & M_{12} & M_{13}\\
M_{21} & M_{22} & M_{23}\\
M_{31} & M_{32} & M_{33}%
\end{array}
\right)  .
\end{equation}

Note that substituting the expressions (49-51) of $x_{i}$ \cite{Hassoul2} in
the diagonalized invariant operator (48) $\mathcal{O}(t)$ does not lead to
expression (28).

A straightforward evaluation of the product $%
\mathbb{R}
^{-1}\Bbbk%
\mathbb{R}
$ leads to the following elements
\begin{equation}
M_{11}=\frac{1}{2}\left(  \varpi_{1}^{2}+\varpi_{2}^{2}\right)  +\frac{1}%
{4}\left(  K_{13}+K_{23}\right)  +\frac{1}{2}K_{12},
\end{equation}%
\begin{equation}
M_{12}=\frac{\sqrt{3}}{2}\left[  \frac{\lambda_{+}\left(  \varpi_{1}%
^{2}-\varpi_{2}^{2}\right)  }{2}\left(  K_{12}-K_{23}-2\Omega^{2}\right)
+\frac{\lambda_{+}}{4}\left(  K_{23}^{2}-K_{13}^{2}\right)  \right]  ,
\label{2}%
\end{equation}%
\begin{equation}
M_{13}=\frac{\sqrt{3}}{2}\left[  \frac{\lambda_{-}\left(  \varpi_{1}%
^{2}-\varpi_{2}^{2}\right)  }{2}\left(  K_{12}-K_{23}+2\Omega^{2}\right)
+\frac{\lambda_{-}}{4}\left(  K_{23}^{2}-K_{13}^{2}\right)  \right]  ,
\label{((3))}%
\end{equation}%
\begin{align}
M_{21}  &  =\frac{1}{4\sqrt{3}\lambda_{+}\Omega^{2}}\left(  \frac
{K_{23}-K_{12}-2\Omega^{2}}{K_{23}-K_{13}}\right)  \left[  \varpi_{1}%
^{2}+\varpi_{2}^{2}-2\varpi_{3}^{2}+K_{12}-\frac{K_{13}+K_{23}}{2}\right]
\nonumber\\
&  +\frac{1}{4\sqrt{3}\lambda_{+}\Omega^{2}}\left[  \varpi_{2}^{2}-\varpi
_{1}^{2}+\frac{K_{23}-K_{13}}{2}\right]  , \label{(4)}%
\end{align}%
\begin{align}
M_{22}  &  =\frac{1}{4\Omega^{2}}\left[  \frac{\left(  K_{23}-K_{12}\right)
^{2}-4\Omega^{4}}{K_{23}-K_{13}}\right]  \left(  \frac{\varpi_{2}^{2}%
-\varpi_{1}^{2}}{2}\right) \nonumber\\
&  +\frac{1}{4\Omega^{2}}\left[  \frac{\left(  \varpi_{1}^{2}+\varpi_{2}%
^{2}\right)  -K_{12}}{2}\left(  K_{23}-K_{12}+2\Omega^{2}\right)
+\frac{\left(  K_{23}-K_{13}\right)  ^{2}}{4}\right] \nonumber\\
&  -\frac{\left(  K_{23}-K_{12}-2\Omega^{2}\right)  }{4\Omega^{2}}\left[
\frac{K_{23}-K_{12}+2\Omega^{2}}{2}+\varpi_{3}^{2}-\frac{K_{23}+K_{13}}%
{4}\right]  , \label{(5)}%
\end{align}%
\begin{align}
M_{23}  &  =\frac{\lambda_{-}}{4\lambda_{+}\Omega^{2}}\left[  \frac{\left(
K_{23}-K_{12}-2\Omega^{2}\right)  ^{2}}{K_{23}-K_{13}}\right]  \left(
\frac{\varpi_{2}^{2}-\varpi_{1}^{2}}{2}\right) \nonumber\\
&  +\frac{\lambda_{-}}{4\lambda_{+}\Omega^{2}}\left[  \frac{\left(  \varpi
_{1}^{2}+\varpi_{2}^{2}-K_{12}\right)  }{2}\left(  K_{23}-K_{12}-2\Omega
^{2}\right)  +\frac{\left(  K_{23}-K_{13}\right)  ^{2}}{4}\right] \nonumber\\
&  -\frac{\lambda_{-}\left(  K_{23}-K_{12}-2\Omega^{2}\right)  }{4\lambda
_{+}\Omega^{2}}\left[  \frac{K_{23}-K_{12}-2\Omega^{2}}{2}+\varpi_{3}%
^{2}-\frac{K_{23}+K_{13}}{4}\right]  , \label{(6)}%
\end{align}%
\begin{align}
M_{31}  &  =\frac{1}{4\sqrt{3}\lambda_{+}\Omega^{2}}\left(  \frac
{K_{12}-K_{23}-2\Omega^{2}}{K_{23}-K_{13}}\right)  \left[  \varpi_{1}%
^{2}+\varpi_{2}^{2}-2\varpi_{3}^{2}+K_{12}-\frac{K_{13}+K_{23}}{2}\right]
\nonumber\\
&  +\frac{1}{4\sqrt{3}\lambda_{+}\Omega^{2}}\left[  \varpi_{1}^{2}-\varpi
_{2}^{2}+\frac{K_{13}-K_{23}}{2}\right]  , \label{(7)}%
\end{align}%
\begin{align}
M_{32}  &  =\frac{\lambda_{+}}{4\lambda_{-}\Omega^{2}}\left[  \frac{\left(
K_{12}-K_{23}-2\Omega^{2}\right)  ^{2}}{K_{23}-K_{13}}\right]  \left(
\frac{\varpi_{1}^{2}-\varpi_{2}^{2}}{2}\right) \nonumber\\
&  +\frac{\lambda_{+}}{4\lambda_{-}\Omega^{2}}\left[  \frac{\left(  \varpi
_{1}^{2}+\varpi_{2}^{2}-K_{12}\right)  }{2}\left(  K_{12}-K_{23}-2\Omega
^{2}\right)  -\frac{\left(  K_{23}-K_{13}\right)  ^{2}}{4}\right] \nonumber\\
&  -\frac{\lambda_{+}\left(  K_{12}-K_{23}-2\Omega^{2}\right)  }{4\Omega
^{2}\lambda_{-}}\left[  \frac{K_{23}-K_{12}+2\Omega^{2}}{2}+\varpi_{3}%
^{2}-\frac{K_{13}+K_{23}}{4}\right]  ,\text{ \ \ \ } \label{8}%
\end{align}%
\begin{align}
M_{33}  &  =\frac{\lambda_{+}}{4\lambda_{-}\Omega^{2}}\left[  \frac{\left(
K_{12}-K_{23}\right)  ^{2}-4\Omega^{4}}{K_{23}-K_{13}}\right]  \left(
\frac{\varpi_{1}^{2}-\varpi_{2}^{2}}{2}\right) \nonumber\\
&  +\frac{\lambda_{+}}{4\lambda_{-}\Omega^{2}}\left[  \frac{\left(  \varpi
_{1}^{2}+\varpi_{2}^{2}\right)  -K_{12}}{2}\left(  K_{12}-K_{23}+2\Omega
^{2}\right)  -\frac{\left(  K_{23}-K_{13}\right)  ^{2}}{4}\right] \nonumber\\
&  -\frac{\lambda_{+}\left(  K_{12}-K_{23}-2\Omega^{2}\right)  }{4\lambda
_{-}\Omega^{2}}\left[  \frac{K_{23}-K_{12}-2\Omega^{2}}{2}+\varpi_{3}%
^{2}-\frac{K_{23}+K_{13}}{4}\right]  ,\text{ \ \ \ } \label{(9)}%
\end{align}
the formula (35) of \cite{Hassoul2} cannot be obtained unless the parameters
of $\Bbbk$ obey
\begin{equation}
K_{12}=K_{13}=K_{23}\text{ \ and \ }\varpi_{1}^{2}=\varpi_{2}^{2}%
=\varpi_{3\text{ \ }}^{2}, \label{condi}%
\end{equation}
which implies that $\Omega^{2}=0$ and the eigenvalues $\Omega_{i}^{2}$ read%
\begin{align}
\Omega_{1}^{2}  &  =\varpi_{1}^{2}+K_{12},\label{lambda1}\\
\Omega_{2}^{2}  &  =\varpi_{1}^{2}-\frac{K_{12}}{2},\label{lambda2}\\
\Omega_{3}^{2}  &  =\varpi_{1}^{2}-\frac{K_{12}}{2}. \label{lambda3}%
\end{align}

Furthermore, as mentioned in Section 2, it is crucial in the Lewis and
Riesenfeld theory \cite{LR} for the invariant operator to have
time-independent eigenvalues whereas in \cite{Hassoul2} the eigenvalues
$\Omega_{i}^{2}$ are time-dependent. To see this, we calculate the time
derivative of $\Omega_{i}^{2}$ as%
\begin{equation}
\frac{d\Omega_{1}^{2}}{dt}=\frac{d}{dt}\left(  \frac{D_{12}}{\sqrt{m_{1}m_{2}%
}}\right)  +\frac{d}{dt}\left(  \frac{D_{13}}{\sqrt{m_{1}m_{3}}}\right)  ,
\end{equation}%
\begin{equation}
\frac{d\Omega_{2}^{2}}{dt}=-\frac{d}{dt}\left(  \frac{D_{23}}{\sqrt{m_{2}%
m_{3}}}\right)  +\frac{d\Omega^{2}}{dt},
\end{equation}%
\begin{equation}
\frac{d\Omega_{3}^{2}}{dt}=-\frac{d}{dt}\left(  \frac{D_{23}}{\sqrt{m_{2}%
m_{3}}}\right)  -\frac{d\Omega^{2}}{dt},
\end{equation}
one can see that even if $\Omega^{2}=0,$ the parameters $D_{12},$ $D_{13},$
$D_{23}$ and the masses $m_{i}$ are defined as time-dependent. Therefore the
eigenvalues $\Omega_{i}^{2}$ are not time-independent as they are supposed to
be. This is once again a fundamental error that contradicts with the Lewis
Riesenfeld theory.

Finally, we suspect that the authors have calculated the phases (56) in
\cite{Hassoul} ((76) in \cite{Hassoul2}) by taking the invariant operator
instead of the Hamiltonian operator in which the term $X_{1}X_{2}$ has been
omitted. Knowing that the dynamics of the system is ruled by the Hamiltonian
operator and not by the invariant operator. It seems that the authors take the
results of \cite{LR, Hartley, Maamache, Pedrosa} and simply set $\frac{1}%
{\rho^{2}}=m(t)$ as if the invariant operator is the generator of the
dynamics. Apparently, they present a study of coupled systems from a quantum
point of view (this has an analog in the classical theory) and claim to prove
that the solution to the time-dependent Schr\"{o}dinger equation with the
mixed term $X_{1}X_{2}$ in the Hamiltonian can be reduced to the solution of a
time-independent Schr\"{o}dinger equation involving the quantum invariant. We
believe that is incorrect because the coupled terms $X_{1}X_{2}$ in the
Hamiltonian have a contribution and cannot be omitted.

This paper is an opportunity to draw the reader's attention to the references
[14-15] cited in \cite{Hassoul} which contain errors in dealing with
time-dependent systems without taking into account the generating function of
the canonical transformation \cite{Lo, Benamira}.

It is clear from the analysis above that Hassoul et al.'s analytical
expressions (23), (26)-(32), (41)-(42) and (56) in \cite{Hassoul} and
expressions (28), (38)-(45) and (76)-(80) in \cite{Hassoul2} cannot be
correct. Consequently, all the physical conclusions derived from such
equations, are based on wrong analytical layout.


\begin{thebibliography}{99}                                                                                               %


\bibitem {Hassoul}S. Hassoul, S. Menouar, J. R. Choi and R. Sever,
\textquotedblleft Quantum dynamics of a general time-dependent coupled
oscillator\textquotedblright, Mod. Phy. Let. B \textbf{35 (14), }2150230 (2021).

\bibitem {Hassoul2}S. Hassoul, S. Menouar, H. Benseridi, and J. R. Choi,
\textquotedblleft Dynamical Invariant Applied on General Time-Dependent Three
Coupled Nano-Optomechanical Oscillators\textquotedblright, Journal of
Nanomaterials \textbf{, }6903563 (2021).

\bibitem {1}Y. S. Kim, M. E. Noz and S. H. Oh,\textquotedblleft A simple
method for illustrating the difference between the homogeneous and
inhomogeneous Lorentz groups\textquotedblright, Am. J. Phys. \textbf{47}(10)
(1979) 892.

\bibitem {2}D. Han, Y. S. Kim and M. E. Noz, \textquotedblleft
Lorentz-squeezed hadrons and hadronic temperature\textquotedblright, Phys.
Lett. A 144(3) (1989) 111.

\bibitem {3}Y. S. Kim and E. P. Wigner, \textquotedblleft Entropy and Lorentz
transformations\textquotedblright, Phys. Lett. A 147(7) (1990) 343.

\bibitem {4}D. Han, Y. S. Kim and M. E. Noz, \textquotedblleft O(3,3)-like
symmetries of coupled harmonic oscillators\textquotedblright, J. Math. Phys.
\textbf{36}(8) (1995) 3940.

\bibitem {5}M. S. Abdalla, \textquotedblleft Quantum treatment of the
time-dependent coupled oscillators\textquotedblright, J. Phys. A, Math.
Gen.\textbf{ 29}(9) (1996) 1997.

\bibitem {6}F. Benamira and L. Ghechi, \textquotedblleft Path integral for a
pair of time-dependent coupled and driven oscillators,\textquotedblright
Czech. J. Phys, \textbf{53}(9) (2003) 717.

\bibitem {7}S. Zhang, J. R. Choi, C. I. Um and K. H. Yeon, \textquotedblleft
Quantum uncertainties of mesoscopic capacitance coupled
circuit\textquotedblright, Phys. Lett. A 289(4--5) (2001) 257.

\bibitem {8}S. Zhang, J. R. Choi, C. I. Um and K. H. Yeon, \textquotedblleft
Quantum squeezing effect of mesoscopic capacitance inductance resistance
coupled circuit\textquotedblright, Phys. Lett. A 294(5--6) (2002) 319.

\bibitem {9}S. Gr\"{o}blacher, K. Hammerer, M. R. Vanner and M.
Aspelmeyer,\textquotedblleft Observation of strong coupling between a
micromechanical resonator and an optical cavity field\textquotedblright,
Nature 460(7256) (2009) 724.

\bibitem {10}W. G. van der Wiel, Y. V. Nazarov, S. De Franceschi, T. Fujisawa,
J. M. Elzerman, E. W. G. M. Huizeling, S. Tarucha and L. P.
Kouwenhoven,\textquotedblleft Electromagnetic Aharonov-Bohm effect in a
two-dimensional electron gas ring\textquotedblright, Phys. Rev. B 67(3) (2003) 033307.

\bibitem {11}T. A. Kennedy, R. Wagner, B. McCombe and D.
Tsui,\textquotedblleft Frequency-Dependent Cyclotron Effective Masses in Si
Inversion Layers\textquotedblright, Phys. Rev. Lett. \textbf{35}(15) (1975) 1031.

\bibitem {12}G. Sadiek, E. I. Lashin and M. S. Abdalla,\textquotedblleft
Entanglement of a two-qubit system with anisotropic XYZ exchange coupling in a
nonuniform time-dependent external magnetic field\textquotedblright, Physica B
Condens. Matter \textbf{404}(12) (2009) 1719.

\bibitem {13}J. R. Choi,\textquotedblleft Exact quantum theory of
noninteracting electrons with time-dependent effective mass in a
time-dependent magnetic field\textquotedblright, J. Phys. Condens. Matter
\textbf{15}(6) (2003) 823.

\bibitem {14}S. Menouar, M. Maamache and J. R. Choi,\textquotedblleft The
time-dependent coupled oscillator model for the motion of a charged particle
in the presence of a time-varying magnetic field\textquotedblright, Phys. Scr.
\textbf{82}(6) (2010) 065004.

\bibitem {15}S. Menouar, M. Maamache and J. R. Choi,\textquotedblleft An
alternative approach to exact wave functions for time-dependent coupled
oscillator model of charged particle in variable magnetic
field\textquotedblright, Ann. Phys. \textbf{325}(8) (2010) 1708.

\bibitem {16}S. Menouar and J. R. Choi,\textquotedblleft A hybrid approach for
quantizing complicated motion of a charged particle in time-varying magnetic
field\textquotedblright, Ann. Phys. 353 (2015) 307.

\bibitem {17}D. Laroze and R. Rivera,An exact solution for electrons in a
time-dependent magnetic field Phys. Lett. A 355(4--5) (2006) 348.

\bibitem {1'}M. Ebert, A. Volosniev, and H. W. Hammer, \textquotedblleft Two
cold atoms in a time-dependent harmonic trap in one
dimension\textquotedblright,\ Ann. Phys.IF 3.32. \textbf{528} (9) (2016) 693.

\bibitem {6'}E. Gil-Santos, M. Labousse, C. Baker et al., \textquotedblleft
Light-mediated cascaded locking of multiple nano-optomechanical
oscillators,\textquotedblright\ Phys. Rev. Lett, \textbf{118} (6) (2017) 063605.

\bibitem {7'}N. Spethmann, J. Kohler, S. Schreppler, L. Buchmann, and D. M.
Stamper-Kurn, \textquotedblleft Cavity-mediated coupling of mechanical
oscillators limited by quantum back- action,\textquotedblright\ Nat. Phys,
\textbf{12} (1) (2016) 27.

\bibitem {8'}A. G. Litvak and M. D. Tokman, \textquotedblleft
Electromagnetically induced transparency in ensembles of classical
oscillators,\textquotedblright\ Phys. Rev. Lett, \textbf{88 }(9) (2002) 095003.

\bibitem {9'}P. R. Hemmer and M. G. Prentiss, \textquotedblleft
Coupled-pendulum model of the stimulated resonance Raman
effect,\textquotedblright\ JOSA B. Phys, \textbf{5}(8) (1988) 1613.

\bibitem {10'}D. B. Sullivan and J. E. Zimmerman, \textquotedblleft Mechanical
analogs of time dependent Josephson phenomena,\textquotedblright\ Am. J. Phys,
\textbf{39} (12) (1971) 1504.

\bibitem {11'}R. Marx and S. J. Glaser, \textquotedblleft Spins swing like
pendulums do: an exact classical model for TOCSY transfer in systems of three
isotropically coupled spins 1/2,\textquotedblright\ J.Magn. Reson,
\textbf{164} (2) (2003) 338.

\bibitem {12'}A. Merdaci and A. Jellal, \textquotedblleft Entanglement in
three coupled harmonic oscillators,\textquotedblright\ Phys. Lett A,
\textbf{384} (6) (2019) 126.

\bibitem {13'}D. Park, \textquotedblleft Dynamics of entanglement and
uncertainty relation in coupled harmonic oscillator system: exact
results\textquotedblright,\ Quantum Information Processing, \textbf{17} (2018) 147.

\bibitem {14'}D. Park, \textquotedblleft Dynamics of entanglement in three
coupled harmonic oscillator system with arbitrary time-dependent frequency and
coupling constants,\textquotedblright\ Quantum Information Processing,
\textbf{18}(9) (2019) 282.

\bibitem {LR}H. R. Lewis and W. B. Riesenfeld, An Exact Quantum Theory of the
Time-Dependent Harmonic Oscillator and of a Charged Particle in a
Time-Dependent Electromagnetic Field, J. Math. Phys, \textbf{10} (1969) 1458.

\bibitem {Hartley}John G. Hartley and John R. Ray, Ermakov systems and
quantum-mechanical superposition laws, Phys. Rev. A 24 (1981) 2873 (1981).

\bibitem {Maamache}M. Maamache, Ermakov systems, \textquotedblleft exact
solution, and geometrical angles and phases\textquotedblright, Phys. Rev. A 52
(1995) 936.

\bibitem {Pedrosa}I. A. Pedrosa, G. P. Serra and I. Guedes, \textquotedblleft
Wave functions of a time-dependent harmonic oscillator with and without a
singular perturbation\textquotedblright, Phys. Rev. A 56 (1997) 4300.

\bibitem {Lo}C. F. Lo and Y. J. Wong, \textquotedblleft Propagator of two
coupled general driven time-dependent oscillators\textquotedblright\ ,
Europhys. Lett, \textbf{32} (1995) 193.

\bibitem {Benamira}F. Benamira and L. Guechi, Comment on: \textquotedblleft
Propagator of two coupled general driven time-dependent
oscillators\textquotedblright, Europhys. Lett, 60 (2002) 649.
\end{thebibliography}
\end{document}